\documentclass[12pt]{amsart}

\newcommand{\be}{\begin{equation}}
\newcommand{\ee}{\end{equation}}
\newcommand{\bea}{\begin{eqnarray}}
\newcommand{\eea}{\end{eqnarray}}
\newcommand{\beaa}{\begin{eqnarray*}}
\newcommand{\eeaa}{\end{eqnarray*}}

\newcommand{\g}{{\bf g}}

\usepackage{amssymb}
\usepackage{pstricks}
\newcommand{\one}{\begin{pspicture}(0,.1)(.5,.6)
\psline(0,.25)(.5,.25)
\end{pspicture} }
\newcommand{\ons}{\begin{pspicture}(0,.1)(.5,.6)
\psline(0,.25)(.5,.25)
\pscircle[fillstyle=solid](.25,.25){.08}
\end{pspicture} }
\newcommand{\onul}{\begin{pspicture}(0,.1)(.25,.6)
\psline(0,0)(.08,.07)
\psline(.08,.07)(.13,.14)
\psline(.13,.14)(.15,.21)
\psline(.15,.21)(.15,.29)
\psline(.15,.29)(.13,.36)
\psline(.13,.36)(.08,.43)
\psline(.08,.43)(0,.5)
\end{pspicture} }
\newcommand{\onur}{\begin{pspicture}(.25,.1)(.5,.6)
\psline(.5,0)(.42,.07)
\psline(.42,.07)(.37,.14)
\psline(.37,.14)(.35,.21)
\psline(.35,.21)(.35,.29)
\psline(.35,.29)(.37,.36)
\psline(.37,.36)(.42,.43)
\psline(.42,.43)(.5,.5)
\end{pspicture} }
\newcommand{\onuls}{\begin{pspicture}(0,.1)(.25,.6)
\psline(0,0)(.08,.07)
\psline(.08,.07)(.13,.14)
\psline(.13,.14)(.15,.21)
\psline(.15,.21)(.15,.29)
\psline(.15,.29)(.13,.36)
\psline(.13,.36)(.08,.43)
\psline(.08,.43)(0,.5)
\pscircle[fillstyle=solid](.15,.25){.08}
\end{pspicture} }

\newcommand{\tws}{\begin{pspicture}(0,.1)(.5,.6)
\psline(0,0)(.5,0)
\psline(0,.5)(.5,.5)
\end{pspicture} }
\newcommand{\twc}{\begin{pspicture}(0,.1)(.5,.6)
\psline(0,0)(.5,.5)
\psline(0,.5)(.5,0)
\end{pspicture} }
\newcommand{\twgh}{\begin{pspicture}(0,.1)(.5,.6)
\psline(0,0)(.1443,.25)
\psline(0,.5)(.1443,.25)
\psline(.5,0)(.3557,.25)
\psline(.5,.5)(.3557,.25)
\psline(.1443,.25)(.3557,.25)
\end{pspicture} }
\newcommand{\twgv}{\begin{pspicture}(0,.1)(.5,.6)
\psline(0,0)(.25,.1443)
\psline(.5,0)(.25,.1443)
\psline(0,.5)(.25,.3557)
\psline(.5,.5)(.25,.3557)
\psline(.25,.1443)(.25,.3557)
\end{pspicture} }

\newcommand{\twu}{\onul \onur}
\newcommand{\twssu}{\begin{pspicture}(0,.1)(.5,.6)
\psline(0,0)(.5,0)
\psline(0,.5)(.5,.5)
\pscircle[fillstyle=solid](.25,.5){.08}
\end{pspicture} }

\newcommand{\twssb}{\begin{pspicture}(0,.1)(.5,.6)
\psline(0,0)(.5,0)
\psline(0,.5)(.5,.5)
\pscircle[fillstyle=solid](.25,0){.08}
\pscircle[fillstyle=solid](.25,.5){.08}
\end{pspicture} }

\newcommand{\trs}{\begin{pspicture}(0,.1)(.5,.6)
\pscircle(.25,.25){.25}
\pscircle[fillstyle=solid](.423,.423){.08}
\end{pspicture} }
\newcommand{\tril}{\begin{pspicture}(0,.1)(.4,.6)
\psline(0,0)(.1443,.25)
\psline(0,.5)(.1443,.25)
\psline(.1443,.25)(.4,.25)
\end{pspicture} }
\newcommand{\trir}{\begin{pspicture}(0,.1)(.4,.6)
\psline(.4,0)(.2557,.25)
\psline(.4,.5)(.2557,.25)
\psline(.2557,.25)(.0,.25)
\end{pspicture} }
\newcommand{\trils}{\begin{pspicture}(0,.1)(.4,.6)
\psline(0,0)(.1443,.25)
\psline(0,.5)(.1443,.25)
\psline(.1443,.25)(.4,.25)
\pscircle[fillstyle=solid](.272,.25){.08}
\pscircle[fillstyle=solid](.0721,.375){.08}
\pscircle[fillstyle=solid](.0721,.125){.08}
\end{pspicture} }

\advance\textheight by \topskip
\makeatletter

\@addtoreset{equation}{section}

\def\section{\@startsection {section}{1}{\z@}{-3.5ex plus -1ex minus
 -.2ex}{2.3ex plus .2ex}{\large\bf\centering}}
\def\subsection{\@startsection{subsection}{2}{\z@}{-3.25ex plus%
 -1ex minus -.2ex}{1.5ex plus .2ex}{\bf}}
\def\subsubsection{\@startsection{subsubsection}{3}{\z@}{-3.25ex plus%
 -1ex minus -.2ex}{1.5ex plus .2ex}{\sl}}
\makeatother

\begin{document}

\baselineskip 17pt
\parindent 10pt
\parskip 9pt

\begin{flushright}
hep-th/0308082\\[3mm]
\end{flushright}
\vspace{1cm}
\begin{center}
{\Large {\bf Bulk and boundary $g_2$ factorized $S$-matrices}}\\
\vspace{1cm} {\large N. J. MacKay\footnote{email: {\tt
nm15@york.ac.uk}} and B. J. Short}
\\
\vspace{3mm} {\em Department of Mathematics,\\ University of York,
\\York YO10 5DD, U.K.}
\end{center}

\vskip 0.2in
 \centerline{\small\bf ABSTRACT}
\centerline{
\parbox[t]{5in}{\small
\noindent We investigate the $g_2$-invariant bulk (1+1D,
factorized) $S$-matrix constructed by Ogievetsky, using the
bootstrap on the three-point coupling of the vector multiplet to
constrain its CDD ambiguity. We then construct the corresponding
 boundary $S$-matrix, demonstrating it to be consistent
with $Y(g_2,a_1\!\times a_1)$ symmetry. }}

\vspace{0.2in}
\section{Introduction}

As a preliminary step in the investigation using tensor methods of
$1+1$-dimensional factorized $S$-matrices with exceptional $\g$
(and Yangian $Y(\g)$) invariance, we investigate the case
$\g=g_2$. The factorized $S$-matrix for the seven-dimensional
representation ${\bf 7}$ of $g_2$ was constructed by Ogievetsky
\cite{ogiev86}, and we use this to construct the $g_2\times
g_2$-invariant $S$-matrix, applicable in the principal chiral
model (PCM). We may choose the $S$-matrix to have a bootstrap pole
for the self-coupling of the ${\bf 7}$ multiplet, and the
bootstrap applied to this process constrains the CDD factor.

We then investigate the corresponding solutions of the boundary
Yang-Baxter equations and the boundary $S$-matrices for the $g_2$
PCM --- that is, the extension to $g_2$ of the calculations
carried out for classical $\g$ in \cite{macka01}. The spectral
decomposition is precisely that expected from the
$Y(g_2,a_1\!\times a_1)$ symmetry \cite{macka02}.

The method used is the diagrammatic technique of Cvitanovic
\cite{cvit}. We denote the cubic antisymmetric invariant of $g_2$
as $\tril$, then construct the ${\bf 7}$ of $g_2$ by taking the
defining representation of $SO(7)$ and restricting to those
$\ons\in SO(7)$ such that $\trils=\tril$. Here $\tril$ satisfies
the identities
\be\label{ids}\trir\tril=-6\,\one,\qquad\twgh+\twgv=2\,\twc-\tws-\twu\,,\ee
and it is (sometimes repeated) application of these which is
needed to carry out the calculations. If we take the tensor
product $$ {\bf 7} \otimes {\bf 7} = {\bf 27} \oplus {\bf 14}
\oplus {\bf 7} \oplus {\bf 1}$$ then the projectors onto the
irreducible $g_2$-representations are $$P_{\bf 27}
=\frac{1}{2}\left(\tws+\twc\right)-\frac{1}{7}\twu\,,\quad P_{\bf
14} =\frac{1}{2}\left(\tws-\twc\right)+\frac{1}{6}\twgh\,,$$
$$P_{\bf 7}=-\frac{1}{6}\twgh\,,\quad P_{\bf
1}=\frac{1}{7}\twu\,.$$

\section{The bulk $S$-matrix}

The following $g_2$-invariant $S$-matrix satisfies the Yang-Baxter
equation \cite{ogiev86,ogiev86b} :
$$S_{(1,1)}(\theta)=S(\theta)=\sigma(\theta)\left(P_{\bf 27}
+[2]P_{\bf 14}+ [8]P_{\bf 7}+[2][12]P_{\bf 1}\right),$$ where
$\sigma(\theta)$ is a scalar prefactor and
$$[y]=\frac{\frac{yi\pi}{12}+\theta}{\frac{yi\pi}{12}-\theta}.$$
Imposing $R$-matrix unitarity on the $S$-matrix gives
$\sigma(\theta)\sigma(-\theta)=1$, and imposing
hermitian-analyticity gives
$\sigma(\theta)\!=\!\sigma(-\theta^{\ast})^{\ast}$. We rewrite the
$S$-matrix as
\be\label{SM}S(\theta)=\omega(\theta)\left(\tws-\frac{6\theta}{i\pi}\twc+\frac{2
\theta}{(i\pi\!-\!\theta)}\twu+\frac{\theta}{(\frac{2i\pi}{3}\!-\!
\theta)}\twgh\right),\ee where
$\sigma(\theta)\!=\!(1\!-\!\frac{6\theta}{i\pi})\omega(\theta)$,
and impose crossing symmetry,
\begin{multline*}
\omega(\theta)\left(\tws-\frac{6\theta}{i\pi}\twc+\frac{2\theta}{(i
\pi\!-\!\theta)}\twu+\frac{\theta}{(\frac{2i\pi}{3}\!-\!\theta)}\twgh
\right) \\
=\omega(i\pi\!-\!\theta)\left(\twu-\frac{6(i\pi\!-\!\theta)}{i\pi}\twc
+\frac{2(i\pi\!-\!\theta)}{\theta}\tws+\frac{(i\pi\!-\!\theta)}{(
\theta\!-\!\frac{i\pi}{3})}\twgv\right).
\end{multline*}
Using $\twgv=-\twgh+2\twc-\tws-\twu$ we find that this is
satisfied if
$$\omega(\theta)=\omega(i\pi\!-\!\theta)\frac{(i\pi\!-\!\theta)(\frac{2i
\pi}{3}\!-\!\theta)}{\theta(\frac{i\pi}{3}\!-\!\theta)}
\quad\Leftrightarrow\quad\sigma(\theta)=\sigma(i\pi\!-\!\theta)\frac{(i\pi\!-
\!\theta)(\frac{2i\pi}{3}\!-\!\theta)(\frac{i\pi}{6}\!-\!\theta)}{\theta
(\frac{i\pi}{3}\!-\!\theta)(\theta\!-\!\frac{5i\pi}{6})}.$$ To
solve for $\sigma(\theta)$ we first introduce
$$\mu_a(\theta)=\frac{\Gamma\left(\frac{\theta}{2i\pi}\!+\!\frac{a}{12}
\right)\Gamma\left(\frac{-\theta}{2i\pi}\!+\!\frac{a}{12}\!+\!\frac{1}{2
}\right)}{\Gamma\left(\frac{-\theta}{2i\pi}\!+\!\frac{a}{12}\right)
\Gamma\left(\frac{\theta}{2i\pi}\!+\!\frac{a}{12}\!+\!\frac{1}{2}\right)
},$$ which satisfies $\mu_a(\theta)\mu_a(-\theta)\!=\!1$ and, for
real $a$, $\mu_a(\theta)\!=\!\mu_a(-\theta^{\ast})^{\ast}$.
Further,
$$\frac{\mu_a(\theta)}{\mu_a(i\pi\!-\!\theta)}=\frac{(\frac{ai\pi}{6}\!-
\!\theta)}{(\theta\!-\!i\pi\!+\!\frac{ai\pi}{6})}=\frac{\mu_{6-a}(i\pi\!
-\!\theta)}{\mu_{6-a}(\theta)}.$$

We seek a minimal $S$-matrix, with no poles on the physical strip.
The factor $\mu_a (\theta)$ has simple poles at $\theta\!=\!-2i\pi
n\!-\frac{ai \pi}{6}$, $\theta\!=\!2i\pi
n\!+\!i\pi\!+\!\frac{ai\pi}{6}$ and simple zeroes at
$\theta\!=\!2i\pi n\!+\!\frac{ai\pi}{6}$, $\theta\! =\!-2i\pi
n\!-\!i\pi\!-\!\frac{ai\pi}{6}$ for $n\!=\!0,1,2,\ldots$. Thus, to
cancel the poles in (\ref{SM}), we are led to
$$\sigma(\theta)=\mu_0(-\theta)\mu_1(\theta)\mu_3(\theta)\mu_4(\theta
),$$ so that
$$\sigma(\theta)=\frac{\Gamma\left(\frac{\theta}{2i\pi}\!+\!\frac{1}{
2}\right)\Gamma\left(\frac{-\theta}{2i\pi}\right)\Gamma\left(\frac{-
\theta}{2i\pi}\!+\!\frac{7}{12}\right)\Gamma\left(\frac{\theta}{2i\pi}\!
+\!\frac{1}{12}\right)\Gamma\left(\frac{-\theta}{2i\pi}\!+\!\frac{3}{4}
\right)\Gamma\left(\frac{\theta}{2i\pi}\!+\!\frac{1}{4}\right)\Gamma
\left(\frac{-\theta}{2i\pi}\!+\!\frac{5}{6}\right)\Gamma\left(\frac{
\theta}{2i\pi}\!+\!\frac{1}{3}\right)}{\Gamma\left(\frac{-\theta}{2i\pi}
\!+\!\frac{1}{2}\right)\Gamma\left(\frac{\theta}{2i\pi}\right)\Gamma
\left(\frac{\theta}{2i\pi}\!+\!\frac{7}{12}\right)\Gamma\left(\frac{-
\theta}{2i\pi}\!+\!\frac{1}{12}\right)\Gamma\left(\frac{\theta}{2i\pi}\!
+\!\frac{3}{4}\right)\Gamma\left(\frac{-\theta}{2i\pi}\!+\!\frac{1}{4}
\right)\Gamma\left(\frac{\theta}{2i\pi}\!+\!\frac{5}{6}\right)\Gamma
\left(\frac{-\theta}{2i\pi}\!+\!\frac{1}{3}\right)}$$ (in fact we
may choose plus or minus this -- our choice of the positive sign
will not affect the $S$-matrix). Thus we have established a
minimal $S$-matrix which is $g_2$ invariant.

The $g_2$ PCM $S$-matrix acts on multiplets which are
representations of $g_2 \times g_2$, and is constructed from two
minimal $S$-matrices together with a CDD factor $X(\theta)$:
$$S^{PCM}_{(1,1)}(\theta)=X_{(1,1)}(\theta)\Big({S(\theta)}_L\otimes{S
(\theta)}_R\Big).$$ In order that $S^{PCM}_{( 1,1)}(\theta)$
satisfy $R$-matrix unitarity and crossing-symmetry we require
$$X_{(1,1)}(\theta)X_{(1,1)}(-\theta)=1\qquad{\rm
and}\qquad\frac{X_{(
1,1)}(\theta)}{X_{(1,1)}(i\pi\!-\!\theta)}=1.$$ To construct $X$
we use
$$(y)=(y)_{\theta}=\frac{\sinh(\frac{\theta}{2}\!+\!\frac{yi\pi}{24})}
{\sinh(\frac{\theta}{2}\!-\!\frac{yi\pi}{24})};$$ this satisfies
$$(y)_{\theta}(y)_{-\theta}\!=\!1,\qquad
\frac{(y)_{\theta}}{(y)_{i\pi-\theta}}=(2y)_{2\theta}\qquad{\rm
and} \qquad(y)=(y\!+\!24).$$ The natural choice might be
$X=-(2)(4)(8)(10),$ where we have allowed two ${\bf 7}$s to fuse
(via simple poles with positive residues) to form either a ${\bf
7}$ (at $\theta=2i\pi/3$) or a ${\bf 14 \oplus 1}$ (at
$\theta=i\pi/6$, yielding a multiplet of mass
$2\cos(\pi/12)={1\over 2}(\sqrt{6}+\sqrt{2})$ times the mass of
the ${\bf 7}$). We must then check that the bootstrap equations
are satisfied for the scattering of a ${\bf 7}$ off a fused ${\bf
7}\subset {\bf 7}\otimes {\bf 7}$ (an intricate calculation
requiring much repeated application of (\ref{ids})). The minimal
$S$-matrix is consistent with this, but the CDD factor requires an
extra factor $(6)^2$, and we must have
$$X_{(1,1)}(\theta)=-(2)(4)(6)^2(8)(10).$$ The apparent double
pole at $i\pi/2$ thus introduced is spurious: it is cancelled by a
simple zero in each minimal $S$.

\section{The boundary $S$-matrix}

We now consider the half-line case. Following \cite{macka01}, we
try a minimal boundary $S$-matrix of the form
$$K(\theta)=\frac{\tau(\theta)}{(1\!-\!c\theta)}(\one+c\theta\ons).$$
The conditions of boundary $R$-matrix unitarity and hermitian
analyticity impose the constraints
$$(\ons)^{\dag}=\ons,\quad\ons\ons=\one,\quad c\in
i\mathbb{R},\quad
\tau(\theta)=\tau(-\theta^{\ast})^{\ast}\quad{\rm
and}\quad\tau(\theta )\tau(-\theta)=1.$$ We must also impose
crossing-unitarity,
\begin{multline*}
\!\!\!\!\!\frac{\tau(\frac{i\pi}{2}\!-\!\theta)}{(1\!-\!c(\frac{i\pi}{2}\!-\!
\theta))}\left(\onul+c({\textstyle\frac{i\pi}{2}}\!-\!\theta)\onuls
\right)=
\\\frac{\omega(i\pi\!-\!2\theta)\tau(\frac{i\pi}{2}\!+\!\theta)}
{(1\!-\!c(\frac{i\pi}{2}\!+\!\theta))}\left(\twu-\frac{6(i\pi\!-\!2
\theta)}{i\pi}\twc+\frac{(i\pi\!-\!2\theta)}{\theta}\tws+\frac{(i\pi\!-
\!2\theta)}{(2\theta\!-\!\frac{i\pi}{3})}\twgv\right)
\left(\onul+c({\textstyle\frac{i\pi}{2}}\!+\!\theta)\onuls\right).
\end{multline*}
After applying (\ref{ids}) we find that this implies
$$\frac{\tau(\frac{i\pi}{2}\!-\!\theta)}{\tau(\frac{i\pi}{2}\!+\!\theta)
}=\frac{\omega(i\pi\!-\!2\theta)(1\!-\!c(\frac{i\pi}{2}\!-\!\theta))(
\theta\!-\!\frac{i\pi}{3})}{(1\!-\!c(\frac{i\pi}{2}\!+\!\theta))(2\theta
\!-\!\frac{i\pi}{3})}\left(14+2i\pi
c\trs+\frac{i\pi}{\theta}+4\left(c
\trs\!+\!{\textstyle\frac{6}{i\pi}}\right)\theta\right),$$
$$\frac{\tau(\frac{i\pi}{2}\!-\!\theta)}{\tau(\frac{i\pi}{2}\!+\!\theta)
}=\frac{\omega(i\pi\!-\!2\theta)(1\!-\!c(\frac{i\pi}{2}\!-\!\theta))(
\theta\!-\!\frac{i\pi}{3})(i\pi\!+\!2\theta)}{(1\!-\!c(\frac{i\pi}{2}\!+
\!\theta))(2\theta\!-\!\frac{i\pi}{3})}\left(\frac{-\alpha}{(\theta\!-\!
\frac{i\pi}{3})}+\frac{1}{\theta}\mp\frac{12}{i\pi}\right),$$
together with (for non-trivial $\ons$) $\twc\onuls\!=\!\pm\onuls$
and $\twgh\onuls\!=\!\alpha\onuls$ for some constant $\alpha$.
Comparing the two expressions we find $\alpha\!=\!0$ and
$$c\trs=-\frac{6(1\!\pm\!1)}{i\pi}.$$ However, $\trs\!=\!0$
together with $\ons\ons\!=\!\one$ has no solutions in odd
dimensions (the eigenvalues of such a matrix are  $\pm 1$, an odd
number of which cannot sum to zero). We thus have
$(\ons)^t\!=\!\ons$ and
$$\frac{\tau(\frac{i\pi}{2}\!-\!\theta)}{\tau(\frac{i\pi}{2}\!+\!\theta)
}=\frac{\omega(i\pi\!-\!2\theta)(1\!-\!c(\frac{i\pi}{2}\!-\!\theta))(
\theta\!-\!\frac{i\pi}{3})(i\pi\!+\!2\theta)}{(1\!-\!c(\frac{i\pi}{2}\!+
\!\theta))(2\theta\!-\!\frac{i\pi}{3})}\left(\frac{1}{\theta}-\frac{12}{
i\pi}\right),$$ or
$$\frac{\tau(\frac{i\pi}{2}\!-\!\theta)}{\tau(\frac{i\pi}{2}\!+\!\theta)
}=[6]\left[\frac{12}{ci\pi}\!-\!6\right]\sigma(2\theta),\qquad(\ons)^t=
\ons\qquad{\rm and}\qquad c\trs=-\frac{12}{i\pi}.$$ Last we have
to impose the boundary Yang-Baxter equation (bYBe). After some
algebra we find that this is satisfied if
$$\tril+\twssb\tril=\frac{c i \pi}{12}\,\twgh\twssu\tril.$$Now
using (\ref{ids}) we find
$$\trir\twssu\tril=\ons+\frac{12}{ci\pi}\,\one.$$ Thus, putting
these two results together,
$$\twssb\tril=\frac{ci\pi}{12}\,\tril\ons\quad\Leftrightarrow\quad
\tril=\frac{ci\pi}{12}\,\trils.$$ Consequently we must have
$c\!=\!\pm\frac{12}{i\pi}$, with $\trils\!=\!\pm\tril$ and
$\trs\!=\!\mp 1$.

In summary, we have shown that the conditions of $R$-matrix
unitarity, hermitian analyticity, crossing unitarity and the bYBe
are satisfied by a minimal boundary `$K$'-matrix
$$\frac{\tau(\theta)}{(1\!\mp\!\frac{12\theta}{i\pi})}\Big(\one\pm
{\textstyle\frac{12\theta}{i\pi}}\ons\Big)=\tau(\theta)\Big(P_--[\pm
1] P_+\Big),\qquad\Big(P_{\pm}=\frac{1}{2}(\one\pm\ons)\Big),$$
where
$$(\ons)^{\dag}=\ons,\qquad(\ons)^t=\ons,\qquad\ons\ons=\one,\qquad
\trils=\pm\tril,\qquad\trs=\mp 1,$$
$$\tau(\theta)\tau(-\theta)=1,\qquad\tau(\theta)=\tau(-\theta^{\ast})^{
\ast},\qquad\frac{\tau(\frac{i\pi}{2}\!-\!\theta)}{\tau(\frac{i\pi}{2}\!
+\!\theta)}=[6]\left[\pm 1\!-\!6\right]\sigma(2\theta).$$ In fact,
since $[1]_{\frac{i\pi}{2}-\theta}/[1]_{\frac{i\pi}{2}+\theta}=
[-7]/[-5],$ the choice of sign is redundant -- both choices give
the same minimal $K$-matrix. We can write it as
$$\frac{\tau(\theta)}{(1\!-\!\frac{12\theta}{i\pi})}\Big(I+{\textstyle
\frac{12\theta}{i\pi}}E\Big)=\tau(\theta)\Big(P_--[1]P_+
\Big),\qquad\Big(P_{\pm}=\frac{1}{2}(I\pm E)\Big),$$ where
$E\!=\!QXQ^{-1}$, $Q\!\in\!G_2$,
$X\!=\,$diag$(1,1,1,-1,-1,-1,-1)$. This is clearly a subspace of
the symmetric space $SO(7)/S(O(3)\times O(4))$; in fact we have $$
E\in\frac{G_2}{SU(2)\times SU(2)},$$ the space of quaternionic
subalgebras of the octonions, as may be seen by considering the
action of $G_2$ on a basic triple of octonions \cite{baez}.

The following constraints are imposed on $\tau(\theta)$:
$$\tau(\theta)\tau(-\theta)=1,\qquad\tau(\theta)=\tau(-\theta^{\ast})^{
\ast},\qquad\frac{\tau(\frac{i\pi}{2}\!-\!\theta)}{\tau(\frac{i\pi}{2}\!
+\!\theta)}=[6]\left[-5\right]\sigma(2\theta).$$ To solve these we
note that
$$\frac{\mu_a(\frac{i\pi}{2}\!-\!\theta)}{\mu_a(\frac{i\pi}{2}\!+\!
\theta)}=-[2a\!-\!6],$$ and we define
$$\eta_a(\theta)=\frac{\Gamma\left(\frac{-\theta}{2i\pi}\!+\!\frac{a}{12
}\right)\Gamma\left(\frac{\theta}{2i\pi}\!+\!\frac{a}{12}\!+\!\frac{1}{4
}\right)}{\Gamma\left(\frac{\theta}{2i\pi}\!+\!\frac{a}{12}\right)
\Gamma\left(\frac{-\theta}{2i\pi}\!+\!\frac{a}{12}\!+\!\frac{1}{4}
\right)},\qquad {\rm so \;\,that}\qquad
\frac{\eta_a(\frac{i\pi}{2}\!-\!\theta)}{\eta_a(\frac{i\pi}{2}\!+\!
\theta)}=\mu_{2a-6}(2\theta).$$ This leads us to
$$\tau(\theta)=\mu_{1/2}(\theta)\mu_6(\theta)\eta_{7/2}(\theta)\eta_{
9/2}(\theta)\eta_5(\theta)\eta_6(\theta).$$ The simple poles of
$\eta_a(\theta)$ are at $\theta\!=\!2i\pi n\!+\! \frac{ai\pi}{6}$
and $\theta\!=\!-2i\pi n\!-\!\frac{i\pi}{2}\!-\! \frac{ai\pi}{6}$,
while the simple zeroes are at $\theta\!=\!-2i\pi n
\!-\!\frac{ai\pi}{6}$ and $\theta\!=\!2i\pi
n\!+\!\frac{i\pi}{2}\!+\! \frac{ai\pi}{6}$, and so the $K$-matrix
is minimal.

The final piece we require for the complete PCM $K$-matrix is the
factor $Y_1(\theta)$, which must satisfy
$$\frac{Y_1(\frac{i\pi}{2}\!-\!\theta)}{Y_1(\frac{i\pi}{2}\!+\!\theta)}=
X_{(1,1)}(i\pi\!-\!2\theta)=X_{(1,1)}(2\theta).$$ We make use of
the fact that
$$\frac{(y)_{\frac{i\pi}{2}-\theta}}{(y)_{\frac{i\pi}{2}+\theta}}=(2y)_{
i\pi-2\theta}=(2y\!+\!24)_{i\pi-2\theta}.$$ Thus the most natural
choice is $$Y_1(\theta)=(1)(2)(-9)^2(-8)(-7)(-6).$$ This has a
physical strip simple pole at $\theta\!=\!\frac{i\pi} {12}$ at
which the minimal $K$-matrix projects onto the subspace associated
with $P_+$ (the smaller one, and the $({\bf 3},{\bf 1})$ of
$a_1\times a_1$ as found in \cite{macka02}). The simple pole at
$\theta\!=\!\frac{i\pi}{6}$ corresponds to an on-shell diagram
which is possible precisely when the bulk three-point coupling of
${\bf 7}$s exists.

We should also check the simpler trial solution of \cite{macka01}
for a minimal $K$-matrix, namely $$K(\theta)=\rho(\theta)\ons.$$
Imposing crossing-unitarity gives
$$\frac{\rho(\frac{i\pi}{2}\!-\!\theta)}{\rho(\frac{i\pi}{2}\!+\!\theta
)}\onuls=\omega(i\pi\!-\!2\theta)\left(\frac{4(\theta\!-\!\frac{i\pi}{3}
)}{(2\theta\!-\!\frac{i\pi}{3})}\twu+\frac{4(i\pi\!-\!2\theta)(i\pi\!-\!
3\theta)}{i\pi(2\theta\!-\!\frac{i\pi}{3})}\tws+\frac{(i\pi\!-\!2\theta)
(\theta\!-\!\frac{i\pi}{3})}{\theta(2\theta\!-\!\frac{i\pi}{3})}\twc+
\frac{(i\pi\!-\!2\theta)}{(2\theta\!-\!\frac{i\pi}{3})}\twgh\right)
\onuls,$$ which implies \begin{multline*}
\frac{\rho(\frac{i\pi}{2}\!-\!\theta)}{\rho(\frac{i\pi}{2
}\!+\!\theta)}\onuls\;= \\ \omega(i\pi\!-\!2\theta)
\left(\frac{4(\theta\!-\!\frac{i\pi}{3})}{(2\theta\!-\!\frac{i\pi}{3})}
\onul\trs+\frac{4(i\pi\!-\!2\theta)(i\pi\!-\!3\theta)}{i\pi(2\theta\!-\!
\frac{i\pi}{3})}\onuls+\frac{(i\pi\!-\!2\theta)(\theta\!-\!\frac{i\pi}{3
})}{\theta(2\theta\!-\!\frac{i\pi}{3})}\twc\onuls+\frac{(i\pi\!-\!2
\theta)}{(2\theta\!-\!\frac{i\pi}{3})}\twgh\onuls\right).
\end{multline*}
For non-trivial $\ons$ we must have $\trs\!=\!0$,
$(\ons)^t\!=\!\pm\ons$ and $\twgh\onuls\!=\!\alpha\onuls$. But, as
pointed out earlier, the constraint $\trs\!=\!0$ is inconsistent
with $\ons\ons\!=\!\one$. Thus there are no non-trivial solutions
of this form.

{\bf Acknowledgments:} NJM would like to thank Tony Sudbery for a
helpful discussion of the octonions, and BJS would like to thank
the UK EPSRC for a PhD studentship.

\end{document}